\documentclass[12pt]{iopart}
\usepackage{graphicx}
\usepackage{iopams}  
\usepackage{amsfonts}
\usepackage{appendix}

\begin{document}

\title[Persistent Homology for Medium-Range Order]
{Persistent Homology and Many-Body Atomic Structure for Medium-Range Order in the Glass}
\author{Takenobu Nakamura$^1$, Yasuaki Hiraoka$^2$,
Akihiko Hirata$^1$, Emerson G. Escolar$^3$ and Yasumasa Nishiura$^1$}
\ead{t.nakamura@wpi-aimr.tohoku.ac.jp}
\address{$^1$ WPI-AIMR, Tohoku University, Japan}
\address{$^2$ Institute of Mathematics for Industry, Kyushu University, Japan}
\address{$^3$ Graduate School of Mathematics, Kyushu University, Japan}

\begin{abstract}
Characterization of medium-range order in amorphous materials and its relation to 
short-range order is discussed.
A new topological approach is presented here 
to extract a hierarchical structure of amorphous materials, 
which is robust against small perturbations 
and allows us to distinguish it from periodic or random
configurations. 
The method is called the persistence diagram (PD)
and it introduces scales into many-body atomic structures 
in order to characterize the size and shape.
We first illustrate how perfect
crystalline and random structures are represented in the PDs.
Then, the medium-range order in the amorphous silica is 
characterized by using the PD.
The PD approach reduces the size of the data tremendously
to much smaller geometrical summaries and has a huge potential to be
applied to broader areas including complex molecular liquid, granular
materials, and metallic glasses.
\end{abstract}
\pacs{61.43.Er, 61.43.Fs, 02.40.Re}
\maketitle

\section{Introduction}
Glasses have become increasingly useful and popular in materials
engineering and industry. 
Nevertheless, its microscopic structure is not as clearly understood  
as crystalline solids. 
This is mainly due to the lack of long-range order (LRO). 
Hence the local structures, instead of LRO, have been extensively studied to characterize amorphous structures. 
In particular, short-range order (SRO), 
which describes atomic configurations of nearest neighbors, 
was well understood both experimentally and theoretically. 
However, it becomes clear that the larger-scale structures beyond neighbor
atoms so-called  medium-range order (MRO) are much more important than SRO in glasses. 
The amorphous structures are essentially characterized by possible connection of SRO that can build up continuous atomic configurations throughout the materials without any periodicity 
\cite{Zallen1983,Polk1971,Bernal1964,Finney1970,Susman1991PRB,Elliot1991PRL,Greaves2007AdvPhys,ElliotLongman}.

The intrinsic structural features of MRO can be detected 
in several ways.
For example, the split of
the second peak in the radial distribution function is a sign for amorphous metals, 
whereas the first sharp diffraction peak of the structure factor is a sign 
for covalent amorphous solids \cite{ElliotLongman}. 
Although these signs are evidence for the existence of MRO, 
their geometric origins are not clearly understood 
like the ``periodicity" of LRO or the ``chemical or packing order" 
of SRO. 
This is because MRO is definitely generated by many-body configurations,
and the 2-body distributions such as the radial distribution or the structure 
factor does not properly describe its geometry.

To describe the many-body atomic structure appearing in amorphous solids, 
angle distributions and statistics of topological quantities have been widely used so far. 
The bond-angle and torsion-angle distributions extract the partial information of 3-body and 4-body configurations, respectively \cite{Zallen1983}. 
Even though these variables describe the configurations beyond the nearest neighbor, 
the scope of the scale is restricted within $m(\le 4)$-body configurations. 

On the other hand, the topological quantification 
of amorphous structures such as  the ring statistics
and the Voronoi polyhedral analyses 
have been used to characterize atomic configurations of covalent and metallic amorphous solids, respectively. 
\cite{Zallen1983,Polk1971,King1967,Guttman1990,Carol1990,Hobbs1998, Franzblau1991,Goetzke1991,Yuan2002,Wooten2002,WHZachariasenJACS1932}.
These methods do not restrict the number of atoms to be considered, and 
are useful to classify the variety of many-body atomic structure in certain situations.
However, they require building some geometric models from atomic configurations based on artificial criteria such as a threshold of the bond length.
Furthermore, they do not explicitly extract the length scale.
Hence they are not suitable for studying multi-scale phenomena, which are supposed to be important to determine MRO. Systematic methodologies to 
study many-body atomic structure with metric information are highly desired.

In recent years, persistent homology and its graphical representation, called the persistence diagram (PD), 
have been invented in the applied mathematics community as a systematic method to extract geometric properties embedded in point cloud data \cite{eh,carlsson}. 
By regarding the atomic configuration as a point cloud data, 
we can apply the method for geometric analysis of materials.
Significantly, this method can deal with many-body configurations, 
and also can provide two length-scales, {\it birth} and {\it death scales}, 
which characterize the size of the $n$-dimensional holes. 
Namely, topological properties with metric information can be derived.
Furthermore, PDs can be computed quite efficiently \cite{pers,phat,cgal} 
so that it can handle complex configurations 
with a huge amount of atoms 
obtained by molecular dynamics (MD) simulations.
Based on these observations, the PD approach is expected to be an adequate tool to characterize MRO in a systematic way.

In this paper, we propose a methodology based on PDs to characterize MRO in amorphous solids.  
We first provide a brief introduction of PDs mainly for readers 
not familiar with them.
Then we show two examples to illustrate the geometric meaning of PDs 
by using FCC crystal and a uniform random configuration.
Then, we calculated PDs for the atomic configurations of amorphous silica obtained by MD simulations. 
Our method can be applied to a wide class of amorphous solids and complex liquids
with atomic configuration data. 
Here, we chose amorphous silica, 
since it is one of the standard amorphous solids, and hence, 
is an appropriate model material for comparing our method 
to the other conventional tools such as ring statistics. 
Moreover, the extensivity of the Betti numbers, 
the structural hierarchy of the geometric objects appearing in PDs, 
and the decomposition method into single atomic components 
are also explained. 
From these arguments, we elucidate the geometry of MRO in amorphous silica. 
As a consequence, we conclude that the PD analysis is 
an appropriate method to describe MRO, 
complementary to the existing tools.

\section{Persistence Diagram}
The input for computing persistence diagrams is a pair $\mathcal{A}=(Q,R)$ of an atomic configuration 
$Q=(\vec x_1, \vec x_2,... ,\vec x_N)$ and radius parameters $R=(r_1,r_2,... ,r_N)$ for an $N$-atom system. 
Here, $\vec x_i\in \mathbb{R}^3$ and $r_i$ are the three-dimensional position and the input radius 
of the $i$-th atom, respectively.
Instead of setting a threshold by an artificial criterion and giving bonds between atoms, we introduce a ball $B_i(\alpha)=\{\vec x\in \mathbb{R}^3\mid ||\vec x-\vec x_i||\leq r_i(\alpha)\}$ at $\vec x_i$ with radius $r_i (\alpha)=\sqrt{\alpha+r_i^2}$ for each $i$-th atom parameterized by $\alpha$.
Then, we study persistent topological features in the union of balls $B(\alpha)=\bigcup_{i=1}^NB_i(\alpha)$ by 
changing the variable $\alpha$.
Typically, for a increasing family of $\alpha$, 
this is a set of inflating atomic balls with centered at $\vec x_i$.

For each $\alpha$, 
$n$-dimensional holes in $B(\alpha)$ such as clusters, rings, and cavities for $n=0,1$, and $2$, respectively, can be identified by homology \cite{hatcher}.
Then, for each $n$-dimensional hole $c_{k}$, we can uniquely assign values $\alpha=b_{k}$ and $\alpha=d_{k}$ 
at which $c_{k}$ first appears and disappears, respectively. 
These values $b_{k}$ and $d_{k}$ are called the birth and death scales 
of the hole $c_{k}$.
The collection
\begin{equation}
D_n(\mathcal{A})\equiv\{(b_{k}, d_{k})\in\mathbb{R}^2\mid k=1,2,\dots \}
\end{equation}
of all these birth and death scales $(b_{k}, d_{k})$ 
is called the $n$-dimensional persistence diagram of $\mathcal{A}$. 
Here, we remark that the minimum value $\alpha_{\rm min}$ of $\alpha$ can be negative, i.e., $\alpha_{\rm min} =-{\rm min}\{r_1^2,r_2^2,... ,r_N^2 \}$,
and the dimension of $\alpha$ (therefore $b_k$ and $d_k$) 
is length squared.

The PD is a two-dimensional scatter plot. It follows from $b_k< d_k$ 
that the points in the PD appear above the diagonal. 
From the construction, the birth and death points $(b_{k}, d_{k})$ are determined 
by the $m$-body atomic configuration $\vec x_{i_1},\vec x_{i_2} ... ,\vec x_{i_m}$
comprising of the $k$-th hole $c_k$.
Namely, the birth and death scales are represented by 
$b_{k}=b_{k}(\vec x_{i_1}, \vec x_{i_2}, ... , \vec x_{i_m})$ 
and $d_{k}=d_{k}(\vec x_{i_1}, \vec x_{i_2}, ... , \vec x_{i_m})$.
Different from the existing many-body variables such as bond angle 
$\theta_k=\theta_k(\vec x_{i_1},\vec x_{i_2},\vec x_{i_3})$ 
represented by $m=3$, or 
torsion angle 
$\phi_k=\phi_k(\vec x_{i_1},\vec x_{i_2},\vec x_{i_3},\vec x_{i_4})$
represented by $m=4$,
the number $m$ of atoms to be considered
is not restricted for the birth and death scales. 
Therefore $b_k$ and $d_k$ are reduced variables 
of metric properties of $c_{k}$.
We use these variables to describe MRO.

Intuitively, the birth scale $b_{k}$ indicates the maximum neighboring distance in $\vec x_{i_1}, \vec x_{i_2},... ,\vec x_{i_m}$, because $c_k$ appears when the largest bond ($n=1$) or cap ($n=2$) in $c_{k}$  is created. On the other hand, the death scale $d_{k}$  indicates the size of $c_{k}$, because $c_k$ disappears when it is covered up in the
inflated atomic ball model $\bigcup_{i=1}^NB_i(\alpha)$. 

We also introduce the life scale 
$d_{k}-b_{k}$ that represents the robustness of the hole $c_k$ under the change of the variable $\alpha$. 
Then, the points far from the diagonal in the PD have long life scales, meaning that they persist in a wide range of $\alpha$. 
In contrast,
the points close to the diagonal, i.e., small life scales, correspond to holes that are sensitive to $\alpha$ and can be considered noise.

In terms of the characterization of MRO, the life and death scales are important. The former distinguishes the proper geometric objects from the topological noise, whereas the latter represents the sizes of holes. 

Persistent diagrams can be computed efficiently 
using freely available open-source software
\cite{pers,phat}.
For the geometric model of the union of balls structure, we use the alpha shapes module in CGAL \cite{cgal}.
We remark that even though the molecular simulations for bulk systems 
use periodic boundary conditions,
we work with the point cloud data as a set of points 
in $\mathbb{R}^3$ as input to the PDs.
Hence, some differences will appear when we count the holes across 
the boundary, although these effects are negligible 
in a sufficiently large system. 

\section{Typical Examples}

Before analyzing the covalent glass structure, 
we introduce two typical examples.
For simplicity, we only consider monatomic systems.
For a monatomic system, 
the input to the PDs is given by the configuration of its atoms 
and a uniform input radius $r$ for all the atoms.
By the construction of $D_n(\mathcal{A})$,
PDs calculated with a given radius $r$ are the same as ones with $r=0$
after the transformation $(b_k,d_k)\to(b_k+r^2,d_k+r^2)$.
Hence, we choose the input radius to be zero, 
and hence the radius with the scale $\alpha$ is given by $r(\alpha)=\sqrt{\alpha}$.

The first example is a periodic configuration 
as a model of perfect crystal. 
In particular, we choose the FCC crystal. 
In the unit cell, there are 4 atomic sites at 
$(0,0,0), (1/\sqrt{2},1/\sqrt{2},0),(1/\sqrt{2},0,1/\sqrt{2})$ and $(0,1/\sqrt{2},1/\sqrt{2})$ 
and the three primitive lattice vectors are set to be $(\sqrt{2},0,0), (0,\sqrt{2},0)$ 
and $(0,0\sqrt{2})$.
Therefore the number density is $\rho=\sqrt{2}$. 
These lengths are chosen to make the distance between nearest neighbors unity. 
We prepare a cubic cluster of the FCC crystal 
in such a way that
the number of unit cells along each edge is $n_L$, 
the length of each side is $L=\sqrt{2}n_L$, 
and hence, the number of atoms is $N=4n_L^3$.

\begin{figure}[t]
\begin{minipage}{\hsize}
\begin{center}
\includegraphics[width=\hsize]{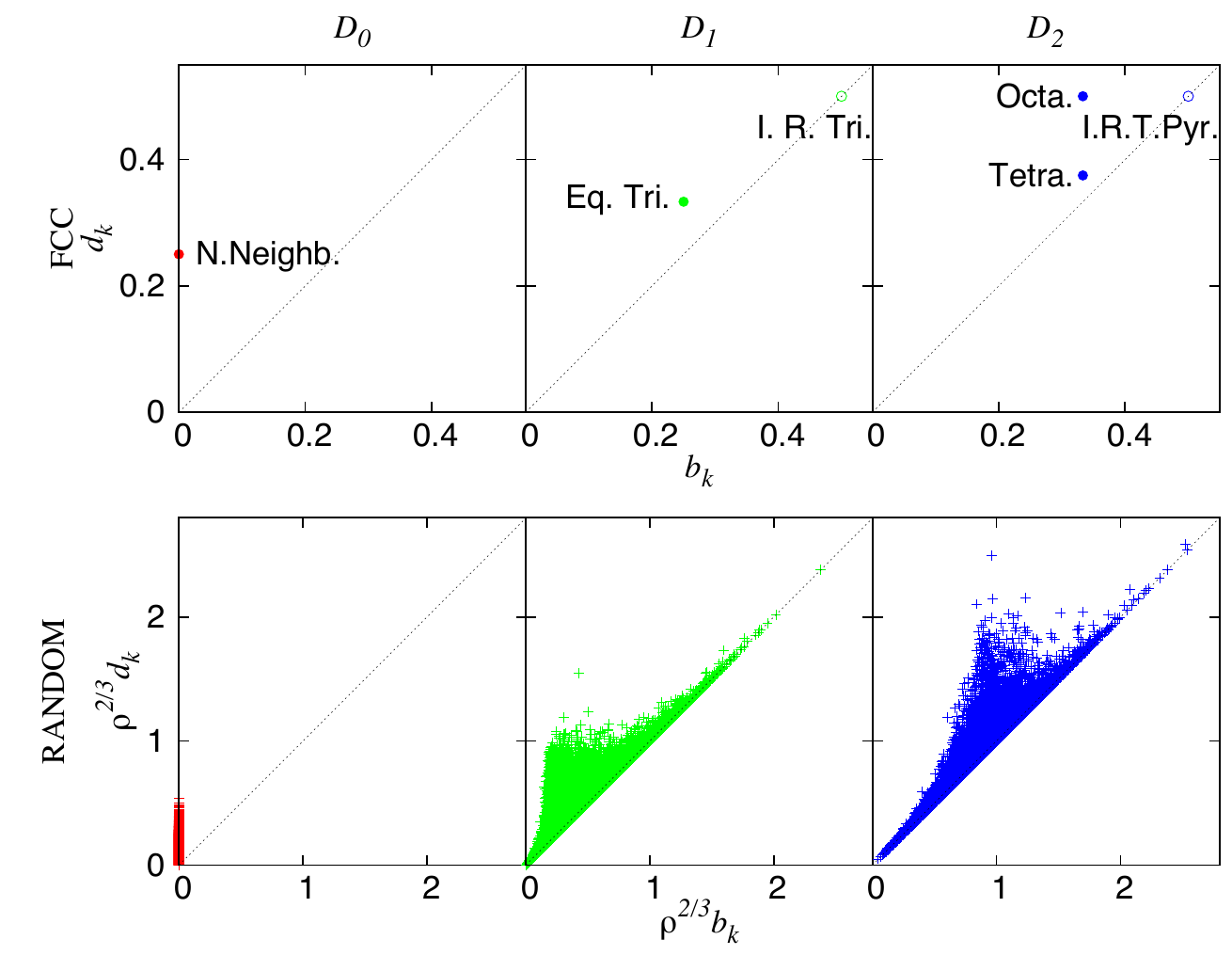}
\caption{
$D_0, D_1$ and $D_2$
are shown in the left, center and right panels respectively.
The top panels correspond to the cubic cluster of the perfect FCC crystal 
with $n_L=10$.
The point at $(0,1/4)$ in $D_0$ represents the contact to the nearest neighbor atoms,
$(1/4,1/3)$ in $D_1$ represents equilateral triangle ring,
$(1/3,3/8)$ and $(1/3,1/2)$ in $D_2$ represent tetrahedron and octahedron cavities respectively.
The bottom panels correspond to a uniform random configuration
composed of $N=27000$ particles scattered in the cubic box $[0,L]^3$.
Each point in the PDs for the perfect FCC crystal has high multiplicity,
whereas PDs for the uniform random configuration have broad distributions.
}
\label{PDs}
\end{center}
\end{minipage}
\end{figure}

The top panels in Fig.~\ref{PDs} show the PDs for the FCC. 
Because of the periodicity, only a few points, each with high multiplicity, 
 are found in each PD. 
The point at $(0,1/4)$ in $D_0$ represents the length to contact the nearest neighbor. 
This is because each inflated atomic ball contacts its nearest neighbor 
atoms at $r_i(\alpha)=1/2$, 
i.e, $\alpha=1/4$. 
In the same way,
the point at $(1/4,1/3)$ in $D_1$ 
represents equilateral triangle rings, 
and the points $(1/3,3/8)$ and $(1/3,1/2)$ in $D_2$ represent 
regular tetrahedral and octahedral cavities, respectively. 
Here, the open circles at the points close to $(1/2,1/2)$ in $D_1$ and $D_2$ represent 
rings that are regular squares or isosceles right triangles
and cavities of isosceles right triangular pyramids, respectively.
These holes are parts of a perfect octahedron.
As the radius of atomic ball becomes large,
these holes are instantly covered right after they 
appeared.

\begin{figure}[t]
\begin{minipage}{\hsize}
\begin{center}
\includegraphics[width=\hsize]{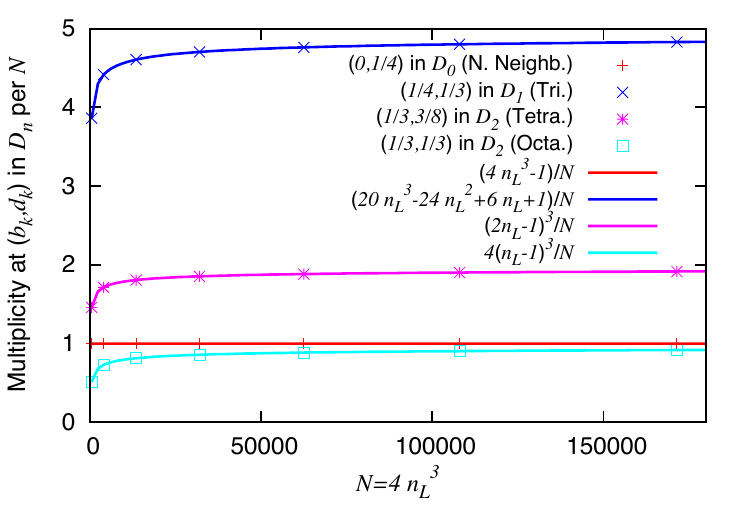}
\caption{
Multiplicities at $(b_k,d_k)$ in $D_n$ per the number of particle $N$ 
for perfect FCC crystal are plotted as a function of $N$.
These values converge to their bulk values like $m(n)/n^3$,
where $m(n)$ is a third-order polynomial.
}
\label{multi_of_N}
\end{center}
\end{minipage}
\end{figure}

The periodicity also leads to high multiplicity.
By a simple geometric consideration,
the multiplicities at $(0,1/4)$ in $D_0$, $(1/4,1/3)$ in $D_1$, and $(1/3,3/8)$ and $(1/3,1/2)$ in $D_2$
are calculated as $m_{\rm N.Neighb}=4n_L^3-1$, 
$m_{\rm Tri.}=20n_L^3-24n_L^2+6n_L+1$, 
$m_{\rm Tetra.}=(2n_L-1)^3$ and $m_{\rm Octa.}=4(n_L-1)^3$ respectively.
The coefficients of $n_L^3$ represent how many clusters,
rings, and cavities exist in the unit cell in the bulk system.
Specifically, in the unit cell,
there are $4$ particles, $8$ tetrahedra and $4$ octahedra.
Every ring in the unit cell is a regular triangle, and is shared 
by a tetrahedron and an octahedron.
Therefore we only need to count the number of rings in the octahedra.
Each octahedron has $8$ triangles on the surface, 
and thus there are $32$ triangle rings in the unit cell.
However, in each octahedron, one ring can be expressed 
as a linear combination  \cite{eh,hatcher} of the other $7$.
Similarly, in each tetrahedron, 
one ring can be expressed as a linear combination of the other $3$. 
Thus, we subtract $4+8$ from $32$, to get a total of $20$ linearly independent rings in each unit cell.
Terms proportional to $n_L^2, n_L$ and the constant come from the boundary effect.
Therefore as the number of atoms becomes large,
the multiplicities per $N$ converge as is seen in Fig.~\ref{multi_of_N}.

In the case of the perfect FCC crystal, 
the points on the diagonal are due to numerical artifacts,
and therefore 
their multiplicities do not scale as $n_L^3$.
If a small perturbation such as thermal fluctuation 
is added to the crystal and resolves the degeneracy,
the points with high multiplicities in the PDs will split 
and distribute with finite width.
Then, the distorted octahedral cavities generate
the secondary rings and cavities close to the diagonal 
and yield the high multiplicities scaled to $n_L^3$,  
as will be shown later (See Fig.~\ref{cryst:finite:temp}).

The second example is a uniform random configuration scattered in a cubic box.
This is a model for the ideal gas. 
In this case, $Q$ is composed of $N$ points 
sampled from the uniform random distribution 
on $[0,L]^3$, and hence the number density of the system is $\rho=N/L^3$.
The bottom panels in Fig.~\ref{PDs} correspond to the PDs 
for the system with $N=27000$. 
The PDs show a broad 2-dimensional distribution 
except for $D_0$ whose birth scales are zero. 

\subsection{$\beta_n(\alpha)$: Betti Number}
\begin{figure}[t]
\begin{minipage}{\hsize}
\begin{center}
\includegraphics[width=0.8\hsize]{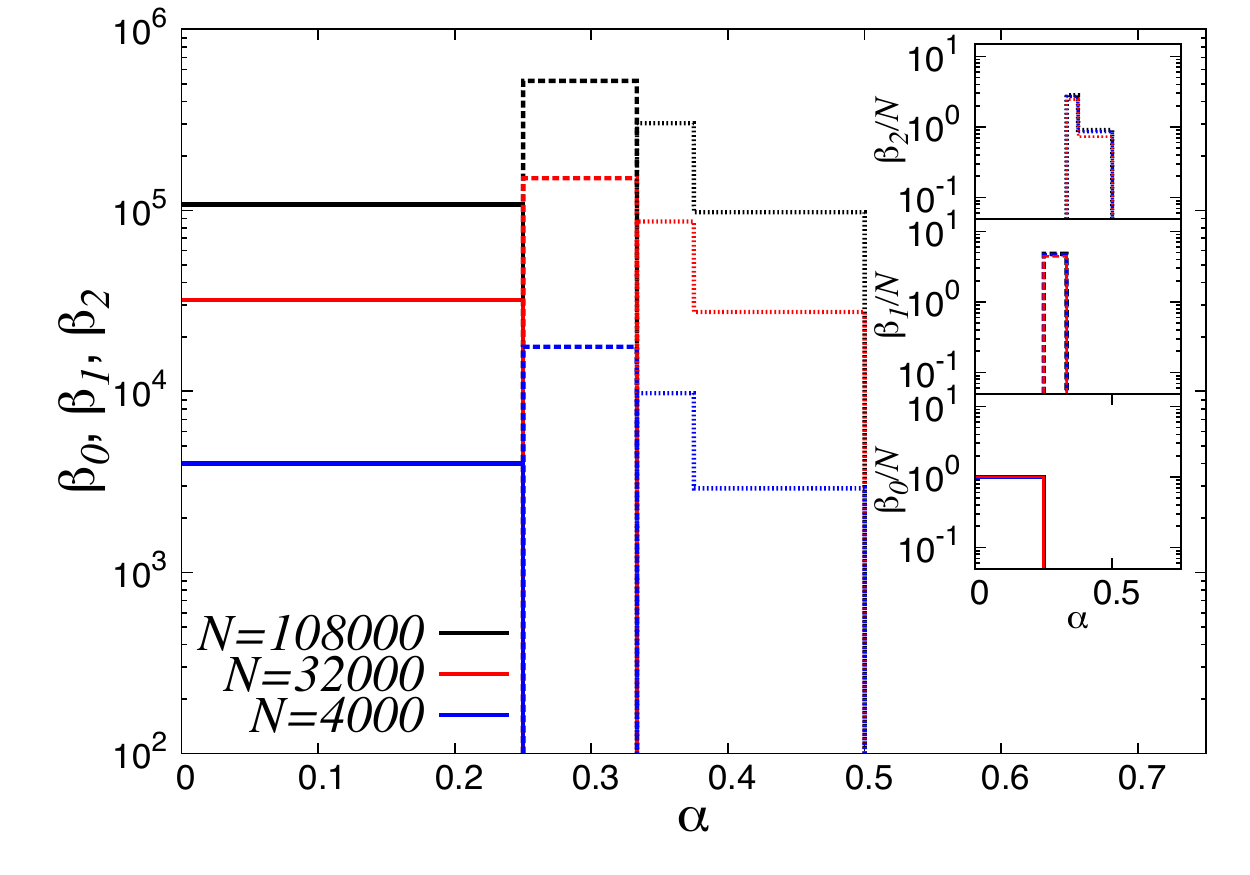}
\includegraphics[width=0.8\hsize]{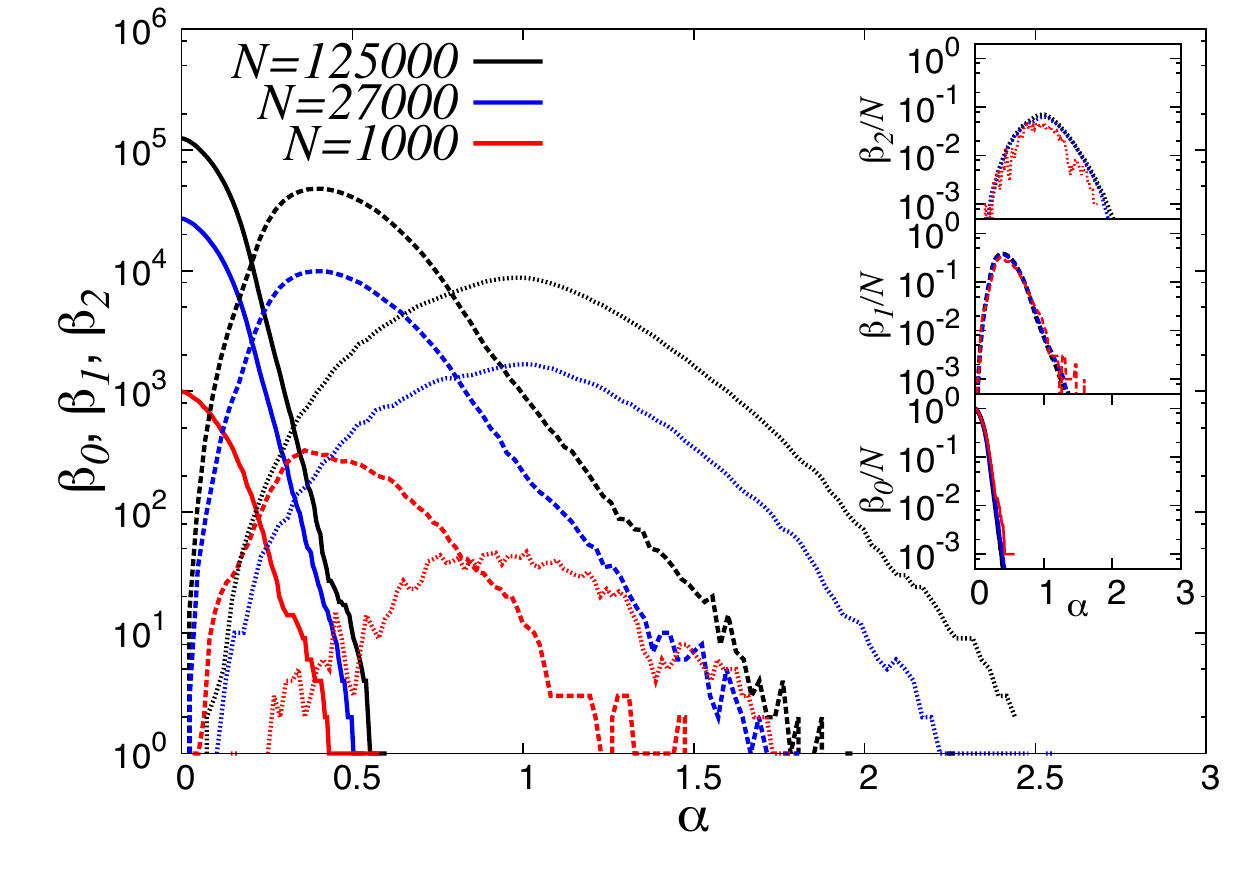}
\caption{Betti numbers $\beta_n (n=0,1,2)$  are plotted against the variable $\alpha$ 
for the perfect FCC crystal with number density $\rho=\sqrt{2}$ (top) 
and the uniform random distribution with fixed number density $\rho=1$ (bottom)
for several $N$.
The solid, dashed, and dots lines correspond to the $\beta_0, \beta_1$ 
and $\beta_2$ respectively.
Insets: Betti numbers per $N$ are plotted as functions of $\alpha$.
}
\label{unscalingN}
\end{center}
\end{minipage}
\end{figure}

As seen in the two examples, 
both the broadness of distribution and the multiplicity 
in the PDs well describe atomic structures of the bulk system.
For further study of the multiplicity, 
we define the Betti numbers $\beta_n(\alpha)$ as
\begin{equation}
\beta_n(\alpha;{\cal A})\equiv 
%\int_{\alpha}^\infty\!\!\!\!\!{\rm d}d\int_{\alpha_{\rm min}}^{\alpha}\!\!\!\!\!{\rm d}b
\int_{\alpha}^\infty{\rm d}d\int_{\alpha_{\rm min}}^{\alpha}{\rm d}b
\sum_{k\in D_n}\delta(b-b_{k})\delta(d-d_{k})\label{def:betti}
\end{equation}
for a given $\alpha$.
In Fig.~\ref{unscalingN}, $\beta_n(\alpha)$ for
$N=4000$, $32000$, and $1715000$ with $\rho=\sqrt{2}$ for the FCC crystal  
(top panel)
and $N=1000$, $27000$, and $125000$ with $\rho=1$
for the uniform random configuration (bottom panel) are plotted.

For the crystal, $\beta_n(\alpha)$ is a piecewise constant function,
whereas $\beta_n(\alpha)$ for the uniform random configuration is 
a single peak function.
The plateau values of $\beta_n$
for the FCC crystal are represented by the multiplicities in $D_n$.
Namely, the plateau values of  $\beta_0,\beta_1$,
the first and second plateau of $\beta_2$ 
are equal to $m_{\rm N.Neighb}, m_{\rm Tri.}, m_{\rm Tetra.}+m_{\rm Octa.}$,
and $m_{\rm Octa.}$ respectively.
Therefore, they are proportional to $N$ for the asymptotically large $N$.
It is noteworthy that $\beta_2(\alpha)$ cannot separate
the number of tetrahedra and octahedra 
even though $D_2$ can.
In this sense, PDs have richer information than the Betti numbers.

\begin{figure}[t]
\begin{minipage}{\hsize}
\begin{center}
\includegraphics[width=\hsize]{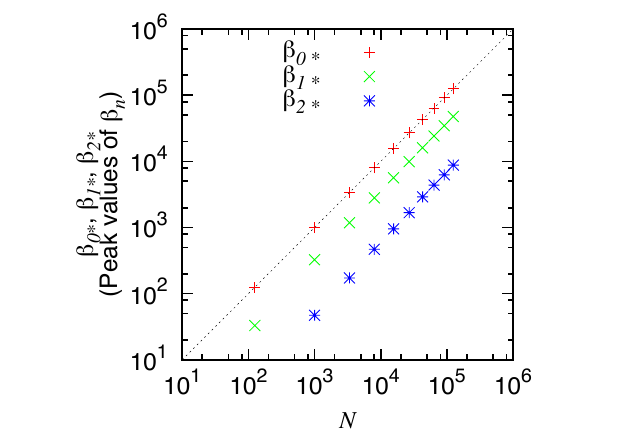}
\caption{The peak values of 
$\beta_n(\alpha)$ for the uniform random configuration are 
plotted against the number of the atoms $N$.
These values are proportional to $N$ in asymptotically large $N$.}
\label{scalingN}
\end{center}
\end{minipage}
\end{figure}

For a uniform random configuration, 
the curves of $\beta_n(\alpha)$ seem to be scaled to $N$
as seen in the insets of the bottom panel in Fig.~\ref{unscalingN}.
To illustrate the scaling, we compare the peak values 
$\beta_{n*}\equiv {\rm max}_{\alpha}\beta_n(\alpha)$ against $N$
and confirm that the peak values $\beta_{n*}$ are 
asymptotically proportional to $N$ (Fig.~\ref{scalingN}).
Therefore, for given $\alpha$, $\beta_n(\alpha)$ is an extensive variable for the system. However, only in the case $n=0$
and for large $\alpha$, this extensively fails because 
$\beta_0(\alpha)=1$ and is independent of the system size.

\subsection{$\xi_n(b,d)$: Normalized Distribution for $D_n$}
Because the Betti numbers are asymptotically proportional to $N$
for the examples above,
it is natural to introduce the two-variable normalized distributions 
for $D_n({\cal A})$
\begin{equation}
\xi_n\left(b,d;{\cal A}\right)\equiv
\frac{\rho^{-4/3}}{N}\sum_{k\in D_n}\delta(b-b_{k})\delta(d-d_{k})
\label{def:xi_n}
\end{equation}
for $n=1$ and $2$.
Here, the factor $\rho^{-4/3}$ is introduced to make $\xi_n$ dimensionless. 
Because $\xi_n$ is a distribution function whose argument is $(b,d)$,
its horizontal and vertical axes are $b$ and $d$,
different from the scatter plot $D_n$ whose axes are $b_k$ and $d_k$.
In the case of $n=0$ for the monatomic system with input radius $r$,
every birth scale takes the same value $b_k=-r$.
Hence, 
we define $\xi_0$ as a function of $d$ by
\begin{equation}
\xi_0\left(d;{\cal A}\right)\equiv
\frac{\rho^{-2/3}}{N}\sum_{k\in D_n}\delta(d-d_{k}).
\label{def:xi_0}
\end{equation}

\begin{figure}[t]
\begin{minipage}{\hsize}
\begin{center}
\includegraphics[width=0.75\hsize]{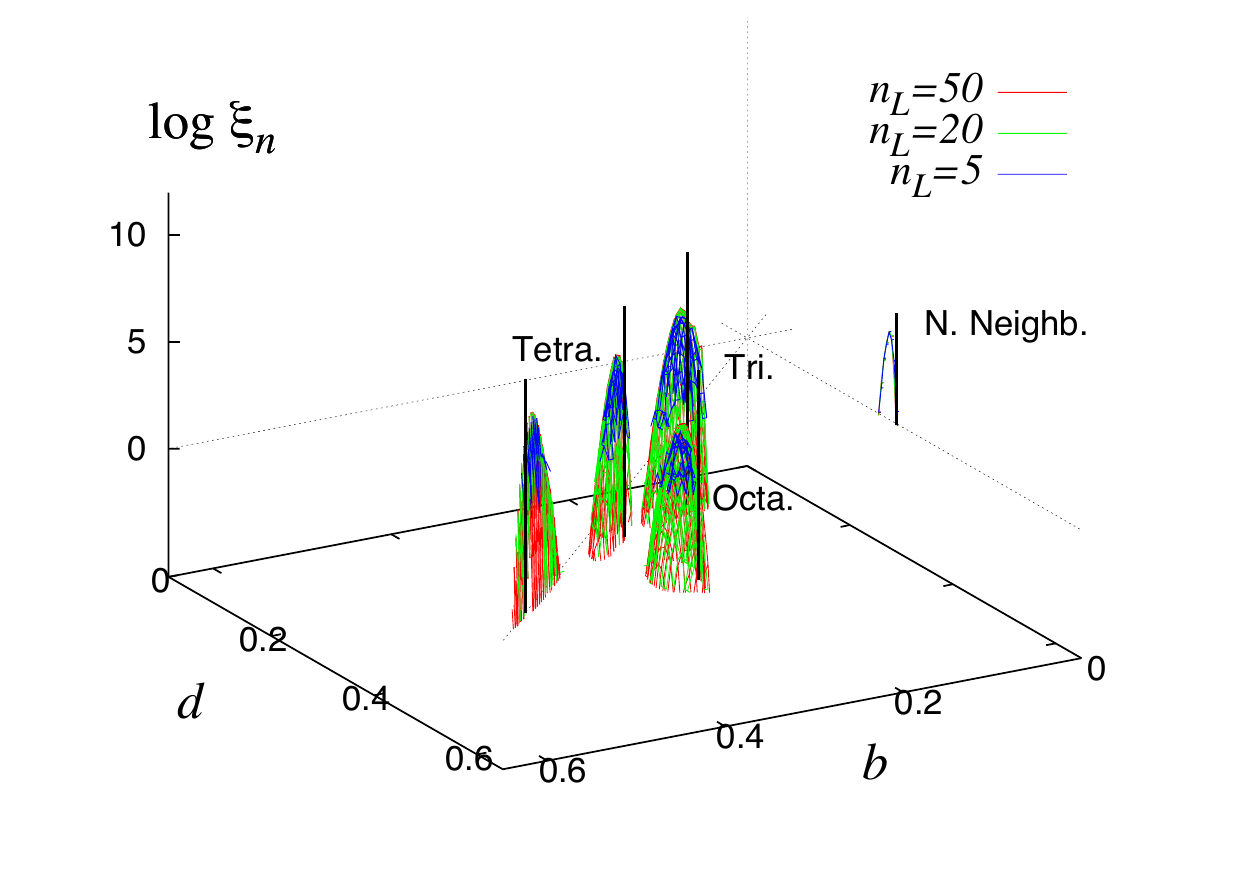}
\includegraphics[width=\hsize]{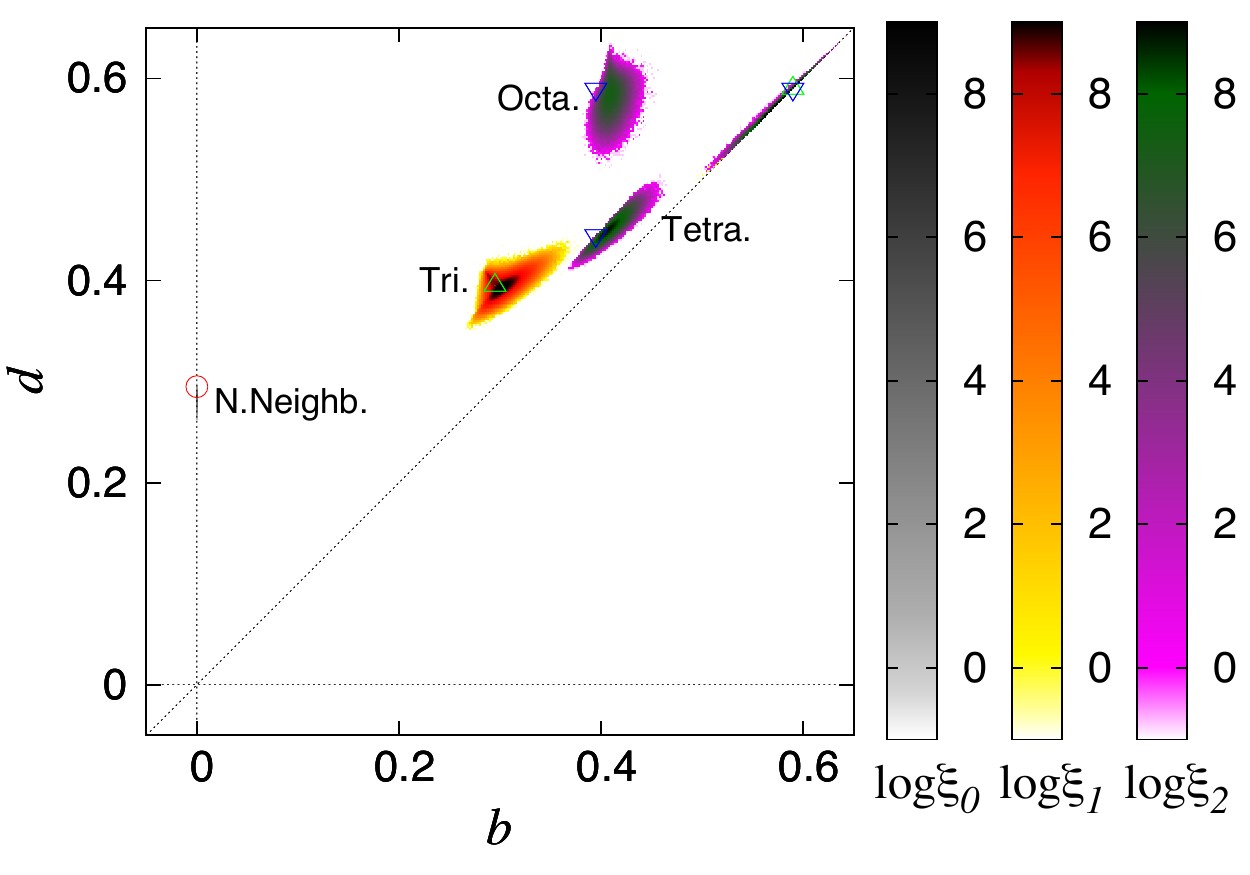}
\caption{
$\xi_n$ for Lennard-Jones FCC crystal 
of temperature $T=0.1$ and pressure $P=1$ in LJ unit.
The convergence of $\xi_n$ for $n_L=50,20,5$ (top panel).
The contour map of $\xi_n$ for $n_L=50$ (bottom panel).
The spikes in top panel and the circle and triangles in bottom panel 
correspond to the configuration obtained by energy minimization, 
and therefore correspond to the configuration of perfect FCC crystal.
}
\label{cryst:finite:temp}
\end{center}
\end{minipage}
\end{figure}

For the FCC crystal, as was seen in Fig.~\ref{multi_of_N},
because the multiplicity for each point asymptotically scales to $N$,
$\xi_n$ converges as $N$ become large.
We also validate this convergence for non-zero temperature crystal
obtained by MD simulation for Lennard-Jones (LJ) particles' system.
The parameters of the potential 
$u(r)=4\epsilon((\sigma/r)^{12}-(\sigma/r)^{6})$ 
are set to be $\epsilon=\sigma=1$.
Starting from the perfect crystal configuration,
$NPT$ (Nos\'e-Hoover-Anderson) simulation 
at temperature $T=0.1$ and pressure $P=1$ in 
LJ unit has been carried out to obtain the equilibrium configuration 
of the FCC crystal.
As a reference, we have also calculated PDs 
for the perfect crystalline configuration
at $T=0,P=1$ obtained by energy minimization.
According to the top panel of Fig.~\ref{cryst:finite:temp},
$\xi_n$ is distributed with non-zero width 
around the spike corresponding to the perfect crystal.
Similar to the perfect crystal,
the distributions for the $\xi_1$ and $\xi_2$ surfaces converge 
asymptotically large $N$.
In the PDs, 
each geometric object forms one isolated island domain over the diagonal,
which is characteristic for the crystalline solid.
In addition, peaks close to the diagonal in $\xi_1$ and $\xi_2$ also 
converge asymptotically large $N$, different from the perfect crystal,
therefore they are no more numerical artifact.
Both rings and cavities in these regions are generated by the octahedra.
This is a typical example, 
where one primary hole generates other secondary holes.
In this case, the octahedral cavity is the primary hole 
and the rings and cavities close to the diagonal are the secondary holes.

\begin{figure}[t]
\begin{minipage}{\hsize}
\begin{center}
\includegraphics[width=0.9\hsize]{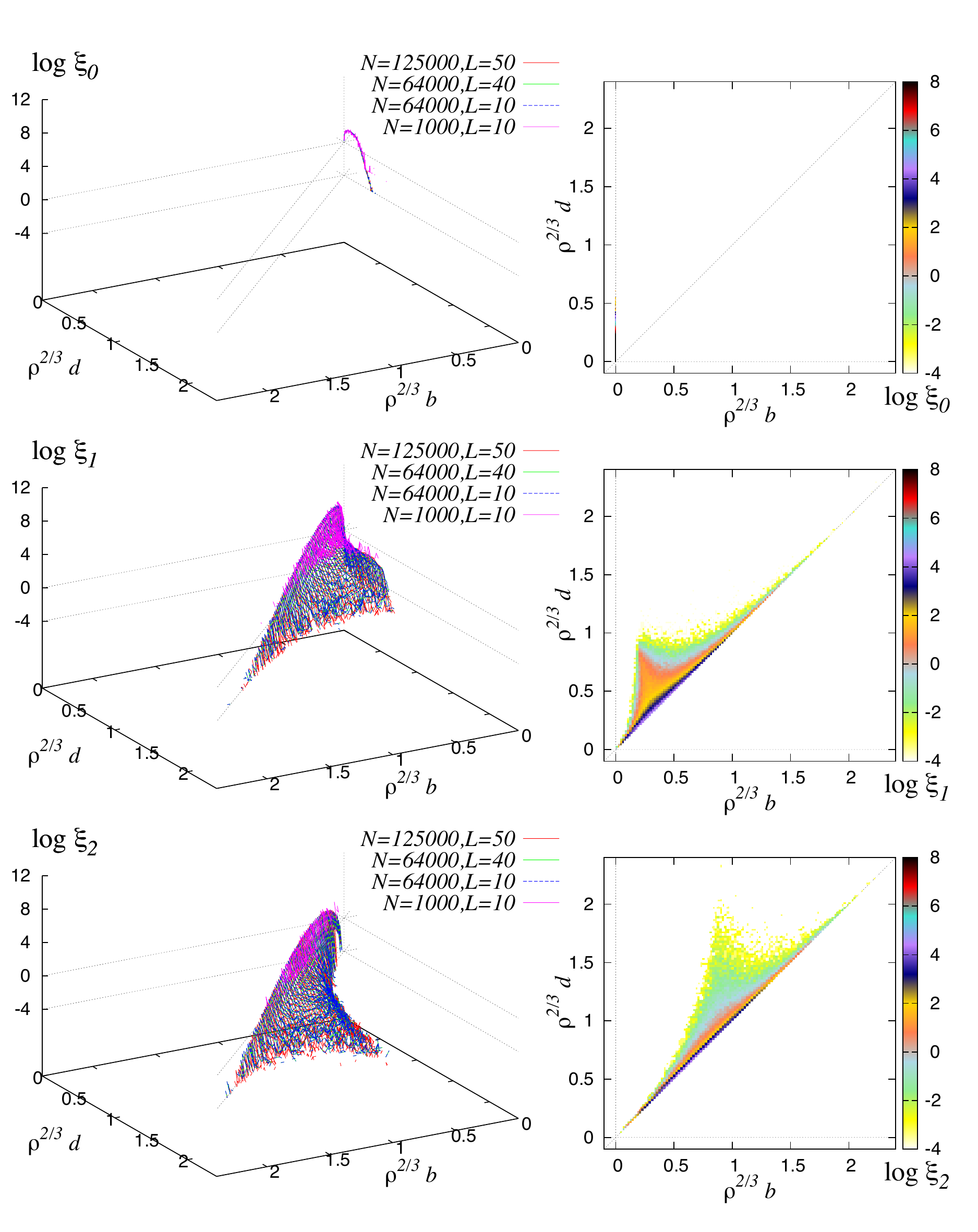}
\caption{The convergence and collapse of 
$\xi_n$ for the uniform random configuration (left panels).
For fixed number density $\rho=N/L^3=1$,
the convergence of $\xi_n$ is observed as $N$ becomes larger
(red, green and magenta surfaces).
For fixed $N=64000$, 
$\xi_n$ for $L=10$ and $40$ (green and blue surfaces) 
collapse as a function of $(\rho^{2/3}b,\rho^{2/3}d)$.
The contour maps of $\xi_n$ (right panels) 
for the uniform random configuration with $N=125000,L=50$
as a function of $(\rho^{2/3}b,\rho^{2/3}d)$.}
\label{scaling:random}
\end{center}
\end{minipage}
\end{figure}

For the uniform random configuration,
$\xi_n(\rho^{2/3}b,\rho^{2/3}d)$ 
is a scaling functions.
Intuitively, the birth and death scales become smaller
as the number density $\rho$ becomes larger.
This effect is cancelled by replacing 
the arguments of $\xi_n$ by $(\rho^{2/3}b,\rho^{2/3}d)$,
as seen in left panels of Fig.~\ref{scaling:random} for $N=6400$, $L=40$ and $10$.
The left panels in Fig.~\ref{scaling:random} 
also show the statistical convergence for $\rho=1$
with $N=1000$, $64000$, and $125000$.
For small $N$, only the points close to the diagonal appear
because holes with short life scales are the majority in the random configuration.
In contrast, a large $N$ is necessary to detect holes with long life scales.
The right panels in Fig.~\ref{scaling:random} show 
the scaling function
for $N=125000$.
The derivation of this scaling function 
is explained in Appendix \ref{app:tilde:xi}.

\begin{figure}[t]
\begin{minipage}{\hsize}
\begin{center}
\includegraphics[width=\hsize]{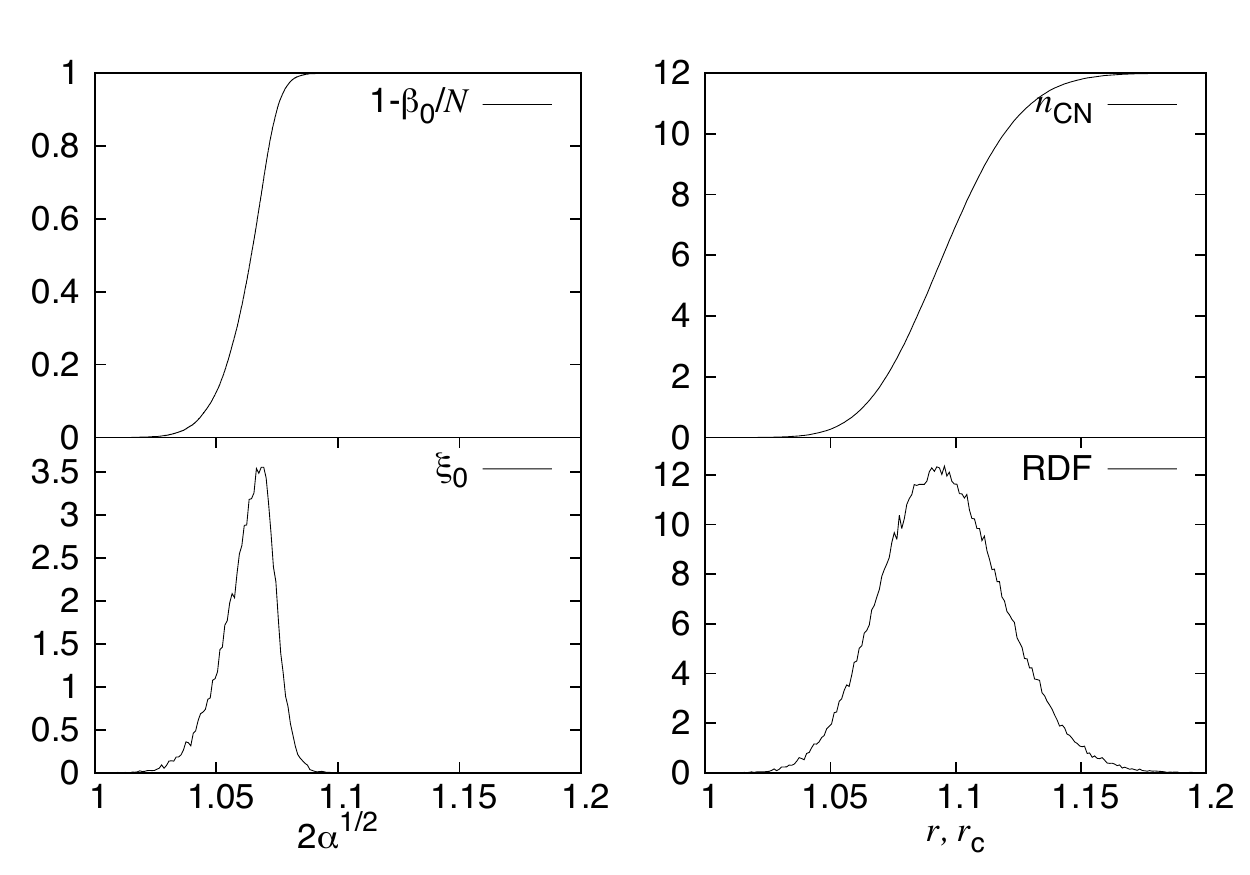}
\caption{Correspondence of 
the relation between radial distribution function (right bottom) and 
coordination number (right top)
and between $\xi_0$ (left bottom) and $1-\beta_0/N$ (left top)
for the FCC crystal of temperature 
$T=0.1$ and pressure $P=1$ in LJ unit.
The length scale in $\xi_0$ and $1-\beta_0/N$ is rescaled 
to $2\alpha^{1/2}$, which corresponds to the distance 
between two atoms in a 2-body distribution.
}
\label{rdfandrcn}
\end{center}
\end{minipage}
\end{figure}

To grasp the meaning of the topological description
using (\ref{def:betti}) and (\ref{def:xi_n}),
we compare them to the coordination number $n_{\rm CN}$ and 
the radial distribution function $g(r)$.
Recall that $n_{\rm CN}$ and $g(r)$ satisfy the following relation:
\begin{equation}
n_{\rm CN}=\int_0^{r_c}dr 4\pi r^2 \rho g(r)
\label{relation:gofr:cn}.
\end{equation}
In Fig.~\ref{rdfandrcn},
we see that $\xi_0$ and $1-\beta_0/N$ show similar behaviors 
to $g(r)$ and $n_{\rm CN}$.
In fact, 
$\beta_0$ and $\xi_0$ satisfy
\[
\frac{\beta_0(\alpha)}{N}=
\rho^{2/3}\int_\alpha^\infty{\mathrm d}d \xi_0(d),
\]
and by using $\beta_0(\alpha_{\rm min})=N$,
we obtain
\begin{equation}
1-\frac{\beta_0(\alpha)}{N}=
\rho^{2/3}\int_{\alpha_{\rm min}}^\alpha{\mathrm d}d \xi_0(d).
\end{equation}
Different from the radial distribution $g(r)$,
$\xi_0(d)$ is determined by many-body atomic configurations. 
Actually, as long as the dimers are counted in the small scale,
they yields the same distribution.
However, as the scale becomes larger, such as
the cluster composed of more than 2 atoms,
these distributions deviate each other.
For $n=1,2$, the relation between $\xi_n$ and $\beta_n$ is given by
\begin{equation}
\frac{\beta_n(\alpha)}{N}=
\rho^{4/3}\int_\alpha^\infty{\mathrm d}d
 \int_{-\infty}^\alpha {\mathrm d}b
 \xi_n(b,d).
\end{equation}

Even though we show the extensivity of Betti numbers $\beta_n$
and the normalization of $\xi_n$ only for the monatomic system, 
these properties are expected to be satisfied even for the multi-component system
as long as the system is macroscopically uniform.

\section{Multi-Component System: Silica Glass}
In this section, we discuss how to apply the PD analysis to multi-component systems, and explain how to get multi-scale geometric information from the PDs. 
We choose silica glasses as examples.
The atomic configurations of silica glasses have been 
obtained by the cooling MD simulation using BKS potentials \cite{prl,BKS,VollmayrPhy.Rev.B.1996}. 
\subsection{Single-Component Analysis}
\begin{figure}[t]
\begin{minipage}{\hsize}
\begin{center}
\includegraphics[width=\hsize]{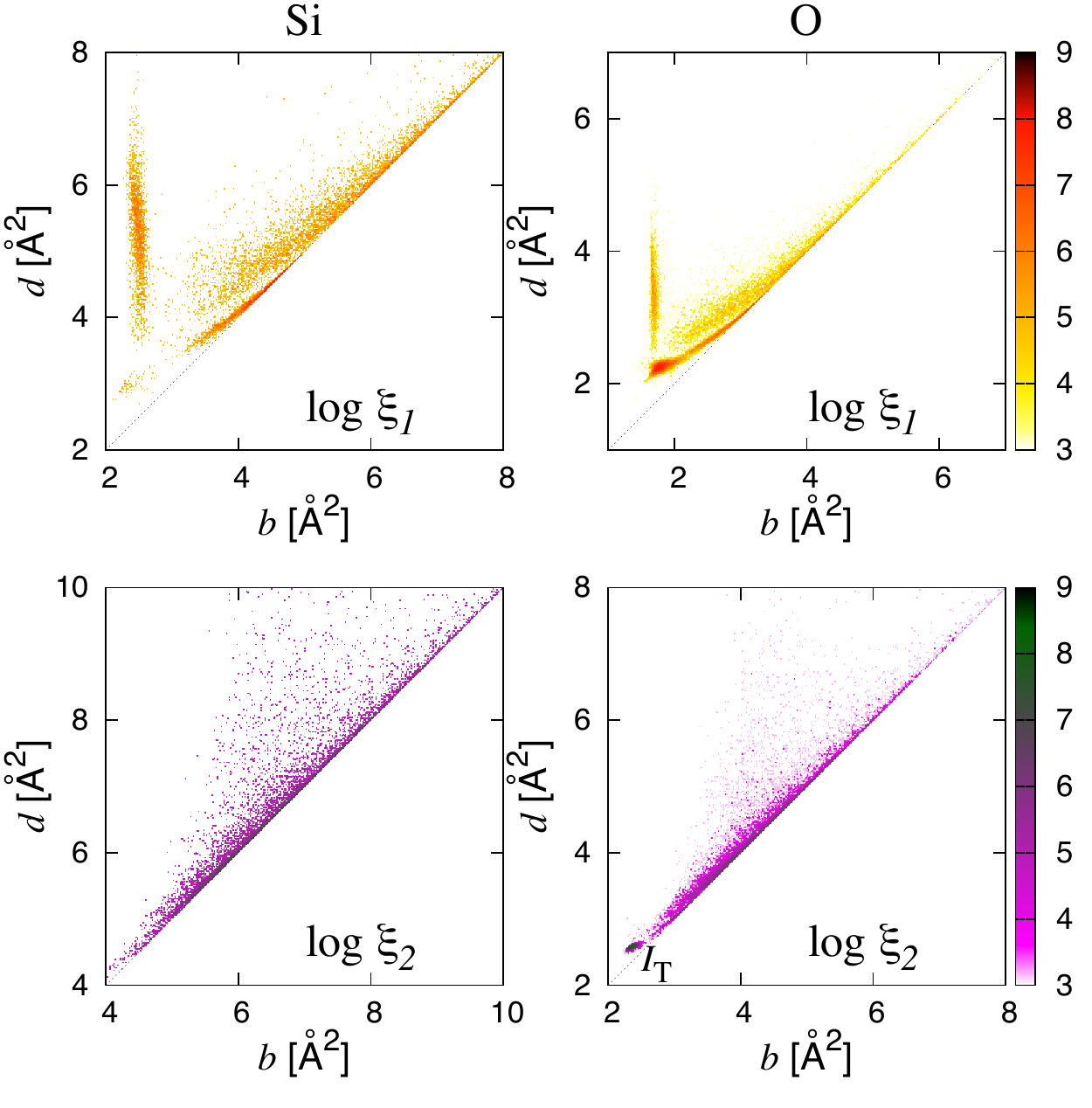}
\caption{PDs for the silicon (left) and oxygen (right) configurations in the amorphous silica. }
\label{single:pd}
\end{center}
\end{minipage}
\end{figure}

In a single-component analysis, the input to the PDs is given by the individual configurations of oxygen and silicon atoms extracted 
from SiO${}_2$ atoms. 
Fig.~\ref{single:pd} shows the PDs of the oxygen and silicon configurations, respectively. 
Similar to the previous section,
we use the uniform input radius $r=0$ for all atoms
by using the invariance under the transformation.
As we see later, this invariance property can be used to determine the atomic compositions of the holes in the multi-component analysis.

We observed that both the birth and death scales of the silicon atoms are larger than those of the oxygen atoms. 
This is because the silicon atoms are distributed more sparsely than the oxygen atoms. 
Furthermore, we found several characteristic domains in the PDs:  
straight lines parallel to the death axis (fixed birth scale) in $D_1$, 
a curve departing from the diagonal in $D_1$,
and a spot domain $I_T$ around $(b,d)=(2.4,2.6)$ in $D_2$ for the oxygen configuration.

Comparing with the PDs for the uniform random distribution, 
these characteristic domains also encode geometric structures 
in the single-component configurations.
For example, the vertical line determined by the fixed birth scale ($b=b_*$) 
in $D_1$ show the existence of a geometric structure in the rings on the line. 
Recall that from the definition, the birth scale measures 
the distance to the neighboring atoms in the ring. 
This means that the rings on the straight line possess the fixed distance $2r(\alpha)=2\sqrt{b_*}$ to the neighboring atoms. 
On the other hand, the diversity of the straight line parallel to the death direction 
means a variety in the sizes of the rings. 
Moreover, we found that a ring on the curve close to the diagonal in $D_1$
is generated by the ring on the vertical line, 
which is similar to 
octahedron in FCC.
We can also show that  the spot domain  $I_{\rm T}$ represents tetrahedra composed of 4 oxygen atoms each.

\subsection{Multi-Component Analysis}

\subsubsection{Existence of Characteristic Curves}
\begin{figure}[t]
\begin{minipage}{\hsize}
\begin{center}
\includegraphics[width=\hsize]{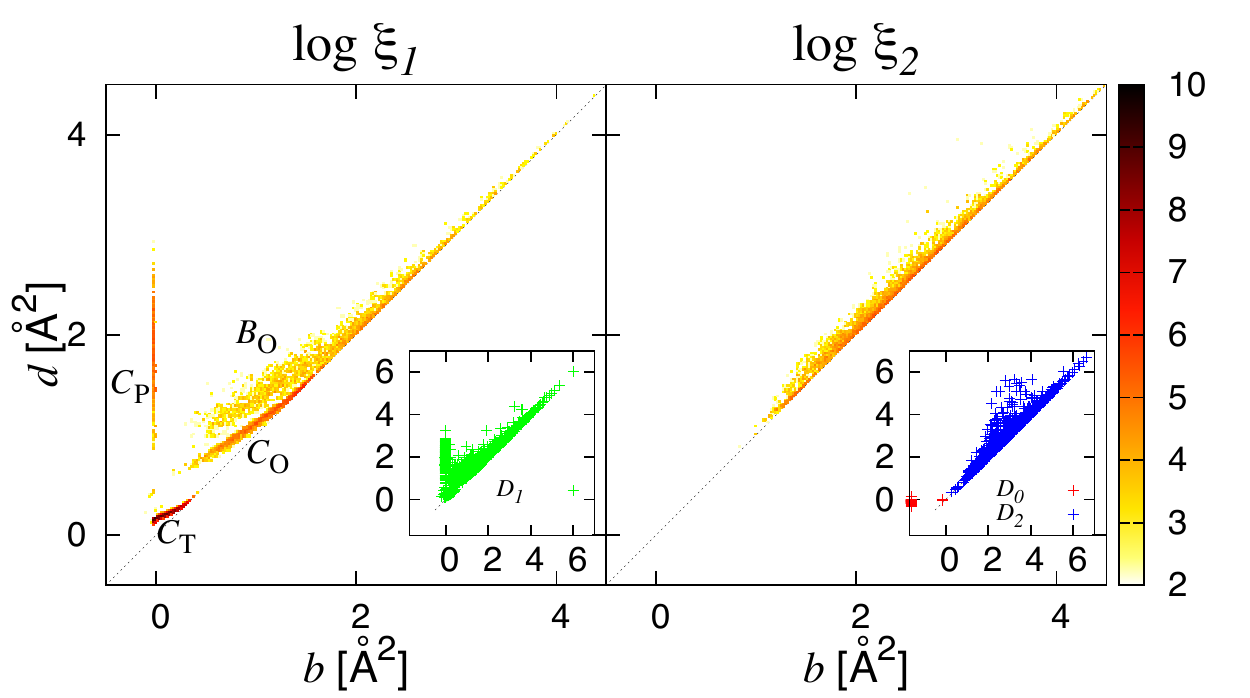}
\caption{
$\xi_1$ (left panel) and $\xi_2$ (right panel) for an amorphous silica.
$D_0, D_1$ and $D_2$  are described in the inset.
The characteristic curves are observed in $\xi_1$.
}
\label{silica:pd}
\end{center}
\end{minipage}
\end{figure}

Different from the single-component analysis, the PDs in this case depend 
on the choice of input radii parameter $R=(r_1,r_2,\dots,r_N)$. 
One of the ways to set $R$ is using partial radial distribution functions of the atomic configuration. Namely, we investigate the first peak positions of pairwise distributions, 
and set the initial radius for each type of atom by solving linear constraints determined by these peak positions. For example, in the atomic configuration of the amorphous silica, we chose the smallest two peaks $d_{\rm SiO}=1.65$\AA~and $d_{\rm OO}=2.55$\AA~(Si-O and O-O pairs, respectively). 
Then, solving the two constraints $r_{\rm Si}+r_{\rm O}=d_{\rm SiO}$ and $2r_{\rm O}=d_{\rm OO}$, we obtain 
the input radii $r_{\rm Si}^*=0.375$\AA~and $r_{\rm O} ^*=1.275$\AA. 
Fig.~\ref{silica:pd} shows the PDs of the amorphous silica for these input radii parameters.

In the previous work \cite{prl}, the PDs of silica glasses and the hierarchical geometric ring structures are studied in detail. We summarize here some of the results which are relevant to the later discussion in this paper. First of all, three characteristic curves, $C_{\rm P}, C_{\rm T}$ and $C_{\rm O}$, 
and one band region $B_{\rm O}$ are found in $D_1$ of Fig.~\ref{silica:pd}. 
The rings on $C_{\rm P}$ have the property 
The rings on $C_{\rm P}$ have the property 
that they generate secondary rings 
when the rings get pinched as the parameter $\alpha$ increases \cite{prl}.
Here P is named after {\em primary} and 
the mechanism is similar to the octahedron cavity in the FCC.
The resulting secondary rings are located on 
$C_{\rm T}, C_{\rm O}$, and $B_{\rm O}$.  

We also note that the cavity distribution in $D_2$ 
for the amorphous silica shows no characteristic regions. 
In particular, the spot domain $I_{\rm T}$ disappears.
This is because the silicon atoms are placed in the interior of the 4-oxygen tetrahedra. 

From the PDs, we can conclude that the ring structures are more essential than the cavities in the amorphous silica.
This is consistent with the fact that existing methods such as ring statistics 
well characterize the amorphous silica. Hereinafter, we only discuss $D_1$.

\subsubsection{Classification of Curves}
\begin{figure}[t]
\begin{minipage}{\hsize}
\begin{center}
\includegraphics[width=0.71\hsize]{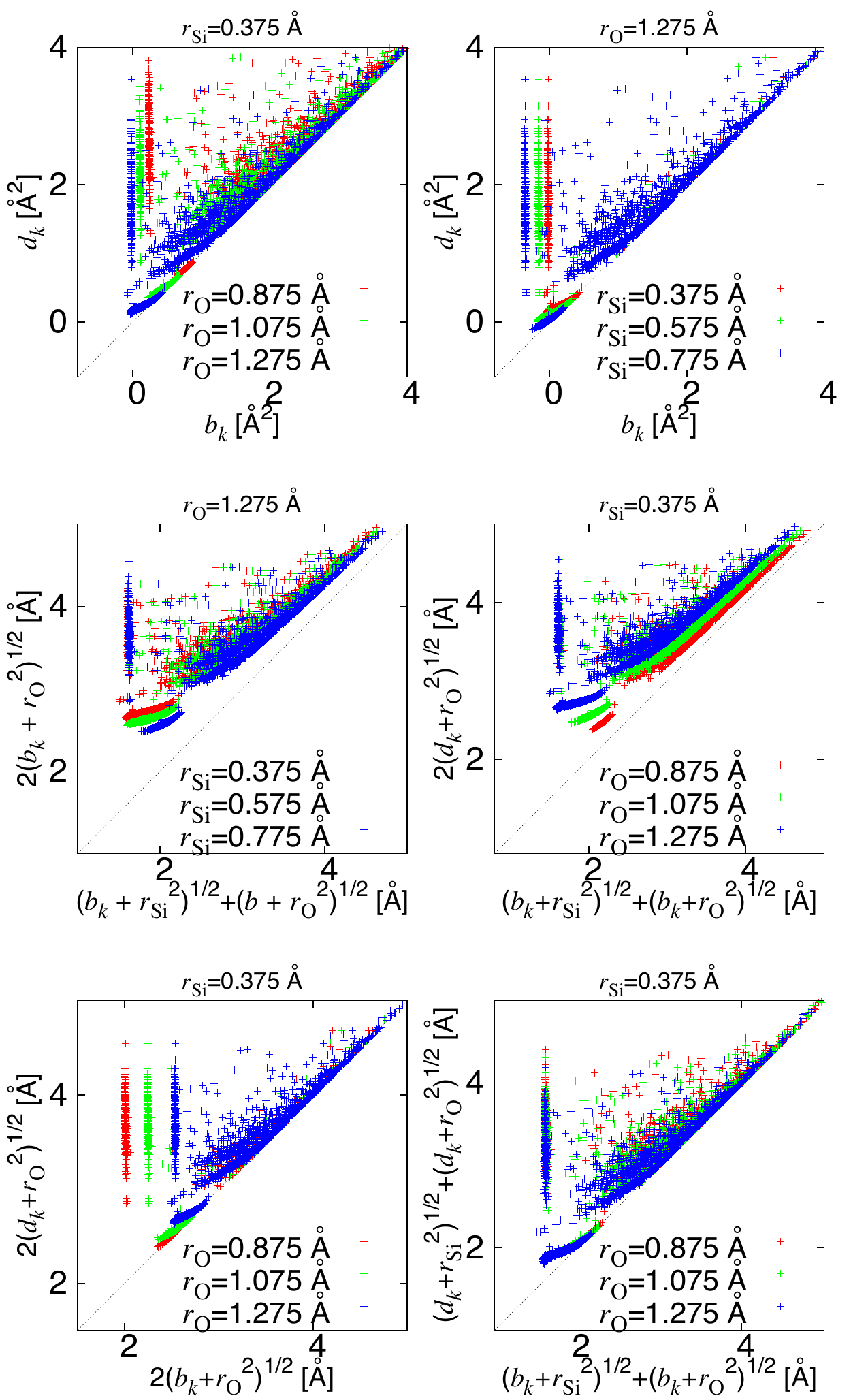}
\caption{The PDs for several input radii are described.
For fixed $r_{\rm Si}=r_{\rm Si}^*=0.375$, 
$r_{\rm O}=r_{\rm O}^*-0.2\AA,r_{\rm O}^*$, and
$r_{\rm O}^*+0.2\AA$ are described in the top left panel.
For fixed $r_{\rm O}=r_{\rm O}^*=1.275$, 
$r_{\rm Si}=r_{\rm Si}^*,r_{\rm Si}^*+0.2\AA$ and
$r_{\rm Si}^*+0.4\AA$ are described in the top right panel.
The overlap of $C_{\rm P}$ is observed 
under the transformation 
$(b_k,d_k)\to (\sqrt{b_k+r_{\rm Si}^2}+\sqrt{b_k+r_{\rm O}^2},2\sqrt{d_k+r_{\rm O}^2})$
in the middle panels.
The collapse of $C_{\rm O}$ and $B_{\rm O}$ region is observed 
under the transformation 
$(b_k,d_k)\to (2\sqrt{b_k+r_{\rm O}^2},2\sqrt{d_k+r_{\rm O}^2})$
in the bottom left panel.
The collapse of $C_{\rm T}$ is observed 
under the transformation 
$(b_k,d_k)\to (\sqrt{b_k+r_{\rm Si}^2}+\sqrt{b_k+r_{\rm O}^2},
\sqrt{d_k+r_{\rm Si}^2}+\sqrt{d_k+r_{\rm O}^2})$
in the bottom right panel.
}
\label{changing:radius:method}
\end{center}
\end{minipage}
\end{figure}

By changing the radius parameter $R$, we can obtain further information such as the composition of the rings. 
In Fig.~\ref{changing:radius:method}, 
several $D_1$ are shown for the different input radii. 
The radii are set to be 
$r_{\rm Si}=r_{\rm Si}^*-0.2{\rm \AA},r_{\rm Si}^*, r_{\rm Si}^*+0.2 {\rm \AA}$, 
and $r_{\rm Si}^*+0.4{\rm \AA}$ for the silicon, 
and $r_{\rm O}=r_{\rm O}^*-0.4{\rm \AA}, r_{\rm O}^*-0.2{\rm \AA}$, and  $r_{\rm O}^*$ for the oxygen. 

As the input radius of oxygen becomes larger, the death and birth scale become smaller 
for fixed input radius of silicon $r_{\rm Si}^*$ 
(the top left panel in Fig.~\ref{changing:radius:method}). 
However, two regions $C_{\rm O}$ and $B_{\rm O}$ 
overlap for the different choice of $r_{\rm Si}$ after the transformation 
$(b_k,d_k)\to (2\sqrt{b_k+r_{\rm O}^2},2\sqrt{d_k+r_{\rm O}^2})$
(the bottom left panel in Fig.~\ref{changing:radius:method}).
As we discussed in the single-component analysis 
(see the right panel in Fig.~\ref{single:pd}.), 
$(b_k+r_{\rm O}^2,d_k+r_{\rm O}^2)$ represents the transformation induced by the input radius $r_{\rm O}$.
Hence, the overlap on the transformed coordinates means that the birth and death scales for the rings on $C_{\rm O}$ and $B_{\rm O}$ are determined by oxygen.

Similarly, we found that $C_{\rm P}$ is invariant to the input radii 
after the transformation $(b_k,d_k)\to 
(\sqrt{b_k+r_{\rm Si}^2}+\sqrt{b_k+r_{\rm O}^2},2\sqrt{d_k+r_{\rm O}^2})$
(the center panels in Fig.~\ref{changing:radius:method}.).
This means the birth scale is determined by the length between silicon and oxygen,
and death scale is determined by the length between oxygen atoms.
For $C_{\rm T}$, we found the invariance via $(b_{k},d_{k})\to
(\sqrt{b_{k}+r_{\rm Si}^2}+\sqrt{b_{k}+r_{\rm O}^2},
\sqrt{b_{k}+r_{\rm Si}^2}+\sqrt{d_{k}+r_{\rm O}^2})$
(the bottom right panel in Fig.~\ref{changing:radius:method}), meaning that the both scales are controlled by the silicon and oxygen atoms. 

Using this procedure, we can classify the regions with respect to the atoms determining the birth and death scales. Furthermore, we can also investigate the compositions of the rings by finding optimal rings \cite{optimal}. For example, it can be shown that the rings on $C_{\rm O}$ and $B_{\rm O}$ consist of only oxygen atoms.

\subsubsection{Geometric Constraints Encoded in Curves}
In contrast to the 2-dimensionally broad distribution of the PDs for the random configuration, 
the curves indicate that there exist geometric relations in the atomic configuration. 
Concretely, the sharp distribution normal to the curve represents a geometric constraint on the rings,
whereas the broad distribution along the tangential direction represents a variety of rings. 

\begin{figure}[t]
\begin{center}
\includegraphics[width=\hsize]{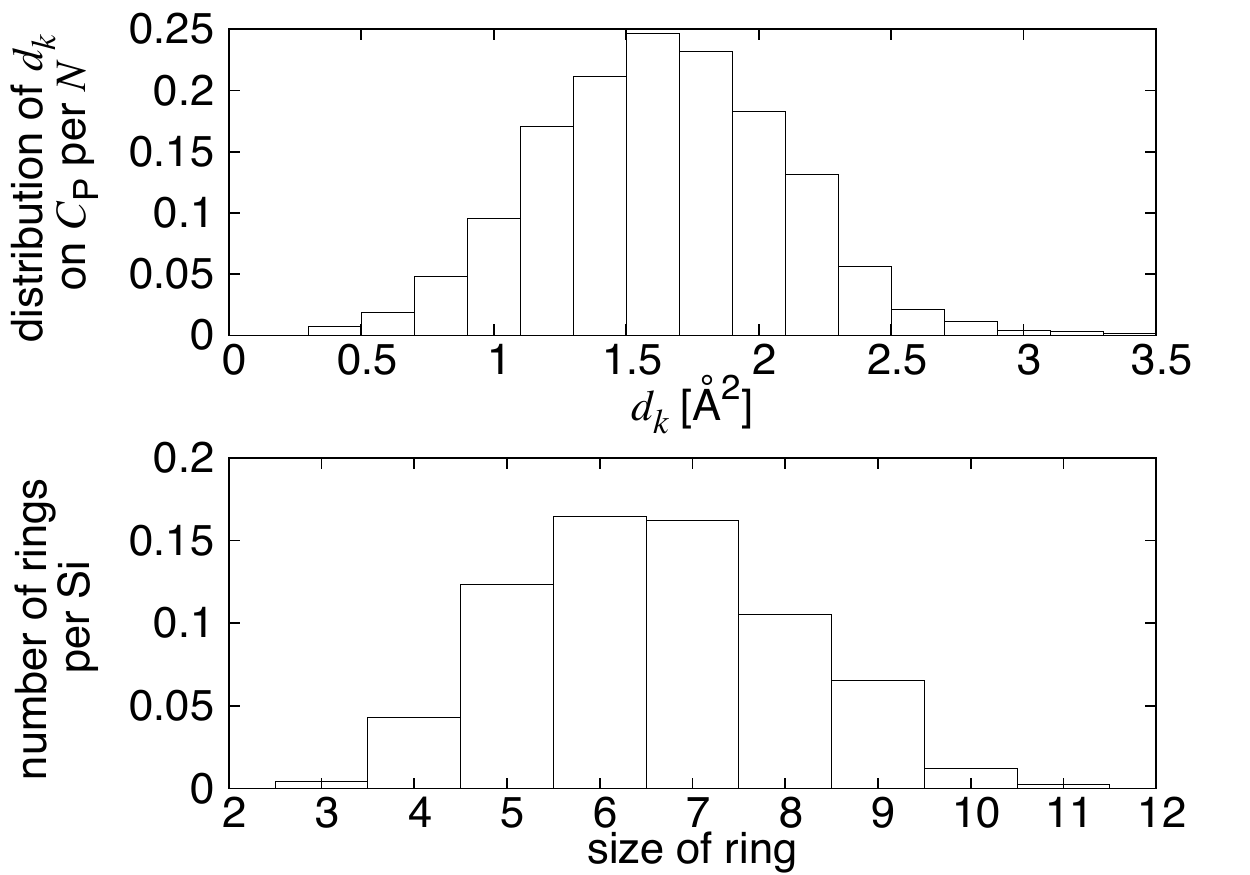}
\caption{
Comparison between the ring statistics (bottom) and the distribution 
of death scales in $C_{\rm P}$ (top).
Both of them represent the size distribution of the rings 
constructed by Si-O network.
}
\label{ringstatistics}
\end{center}
\end{figure}

As shown in Fig.~\ref{silica:pd}, $C_{\rm P}$ lies 
parallel to the death axis on a fixed birth scale. 
It follows from the invariance with respect to the input radii 
after the transformation 
that the birth scale is determined by the bond between silicon and oxygen. 
Thus, we can conclude that the geometric constraint 
in $C_{\rm P}$ means there is a sharp distribution of the Si-O bond length. 
On the other hand, the diversity of the arrangements of the rings 
is mainly controlled by the arrangement of oxygen atoms, 
and this is represented by the diversity along the death axis. 
The death scales in $C_{\rm P}$ 
describe the size of ring constructed by Si-O network,
and hence its diversity corresponds to 
the ring statistics \cite{Zallen1983,ElliotLongman}.
Therefore, as is shown in Fig.~\ref{ringstatistics},
the ring statistics for the shortest path rings \cite{Franzblau1991}
and the death scales in $C_{\rm P}$ show similar distribution.
Even though both of them represent the size distribution of the rings,
only the death scale encodes the metric information.
 
\begin{figure}[t]
\begin{center}
\includegraphics[width=0.7\hsize]{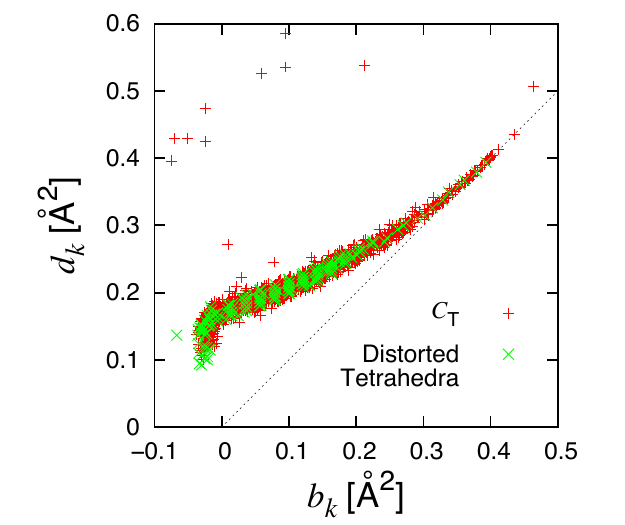}
\caption{
$C_{\rm T}$ (red) and $D_1$ for the distorted tetrahedra (green) are described. 
}
\label{origin:of:T}
\end{center}
\end{figure}

Because the birth and death scales of $C_{\rm T}$ are controlled 
by the bond lengths between silicon and oxygen, 
$C_{\rm T}$ represents short range order. 
Fig.~\ref{origin:of:T} shows both $C_{\rm T}$ and 
the $D_1$ for the distorted tetrahedra. 
Here, the distorted tetrahedra are constructed 
in such a way that tetrahedra in the local arrangement of SiO${}_4$ 
require Si-O and O-O bond lengths to be 
within the first peak of the distributions.
Furthermore, their O-Si-O angles and Si-OOO spherical angles 
are restricted within the 2.5 and 1.5 times the width of the deviation 
around their mean value, respectively, 
determined from the configuration obtained by the MD simulation.
We found that the curve for the tetrahedron deviates from the $C_{\rm T}$  
if we remove one of the restrictions above.
From this observation, the agreement of the two PDs indicates
that the geometric structure represented by the curve $C_{\rm T}$ is a distortion of the SiO${}_4$ tetrahedra.
 
According to Fig.~\ref{silica:pd},
the birth scales of $C_{\rm O}$ and $B_{\rm O}$ are larger 
than the death scales of $C_{\rm T}$. 
This means that they appear after all triangles in SiO${}_4$ tetrahedra 
have been covered by the inflated atomic balls. 
In addition, recall that the rings on $C_{\rm O}$ and $B_{\rm O}$
consist of the oxygen atoms in the primary rings at $C_{\rm P}$. 
Thus, they are smaller than $C_{\rm P}$, representing the order related to the neighboring of tetrahedra, 
the so-called Short Range MRO (SRMRO). 
The geometric constraint corresponding to $C_{\rm O}$ shows the relationship 
between O-O length distribution and O-O-O angle distribution. 
For further details, we refer the reader to the paper \cite{prl}.

Because PDs can encode many-body atomic structure,
we can get a birds-eye view of the geometric hierarchical structure 
among the various kinds of SRO and MRO. 
The key point is that a curve in a PD encodes an order in the atomic configuration. 
Namely, the curve $C_{\rm T}$ represents SRO, whereas
the curves $C_{\rm P}, C_{\rm O}$ and $B_{\rm O}$ show the geometric structure corresponding to MRO in amorphous silica consistent with existing methods. 

From the invariance to the input radii, 
we can convert the death scale $d_{k}$ of each ring to the real space length 
$2\sqrt{d_{k}+r_{\rm O}^2}$, which shows the diameter of the ring. 
Since the death scale for the MRO rings ($C_{\rm P}, C_{\rm O}$ and $B_{\rm O}$) is determined by 
oxygen, this diameter is independent of the choice of the input radii.
The paper \cite{prl} shows that this length scale reproduces the typical length scale of MRO corresponding to the first sharp diffraction peak. 

\section{Concluding Remarks}
Persistent homology is a suitable tool to extract many-body 
atomic structures, and successfully describes geometric features of MRO. 
The topological approach dramatically reduces the degree of freedom 
to represent the many-body structures, 
while keeping metric information and qualitatively important geometric 
features in MRO. 

Several concepts in PD for atomic structures 
are introduced by choosing two extreme cases:
crystalline and random structures.
For the crystalline solid, the periodic structure 
yields a few island supports in $\xi_n$ with high multiplicity, 
and the diagonal region corresponds to the secondary holes 
that represent distortion from the primary holes. 
The shape fluctuation for each hole
is represented as a broadness of the peaks in each island.
For the random configuration, 
there are no isolated supports in $\xi_n$,
and a single broad support is found.
These observations illustrate the power of PDs to express many-body atomic structures.

As an example of network forming glass materials, 
$\xi_1$ for the silica glass shows 
neither island support nor single broad support 
but rather characteristic curves
that imply the existence of MRO.
The primary curve in $\xi_1$ describes the size distribution of rings, 
which is similar to the ring statistics. 
Combining with other curves in $\xi_1$, 
geometric constraint in the MRO is encoded in the PD.
With the aid of length scales encoded in PDs,
a hierarchical and multi-scale many-body atomic structures
are more successfully represented in an integrated manner
compared to the existing methods.
We believe that this method also has a great potential 
to describe the other disordered systems,  such as complex molecular liquids,  packed granular materials, and metallic glasses. 

The analysis by the persistent homology is an integrated method 
incorporating the various existing topological tools. 
The variables $\beta_0$ and $\xi_0$ are the many-body counterparts of
the radial distribution function and the coordination number.
The variable $\xi_1$ in $C_{\rm P}$ corresponds to the ring statistics.
The variable $\xi_2$ might correspond to 
the statistics of the Voronoi index,
although not mentioned in this paper.
Additionally Betti numbers $\beta_n$ shows extensivity
at least for monatomic systems.
It is expected that 
the extensivity is still satisfied for the multi-component system.
By using this behavior,
the phase transition between macroscopically isotropic phases 
such as liquid-liquid transition 
might hopefully be more clearly categorized in the phase diagram.

\appendix

%\addcontentsline{toc}{section}{Appendix} 
\addtocontents{toc}{\protect\setcounter{tocdepth}{-1}}

\section{The Scaling of $\xi_n$ for Uniform Random Configurations}\label{app:tilde:xi}
Suppose that $N$ points are sampled 
from the uniform random distribution
in the $D$-dimensional box $ [0, L]^D$.
Let the $N$-point configuration be denoted as
$Q_{\rm rnd}^{N,L}=(\vec x_1,\vec x_2,...,\vec x_N)$.
Here $\vec x_i$ is a $D$-dimensional position, and $N$ is sufficiently large.
Then, the distribution of the $n$-th persistence diagram 
$D_n(Q_{\rm rnd}^{N,L})$ is given as
\begin{equation}
{\Xi}_n\left(b,d;Q_{\rm rnd}^{N,L}\right)
\equiv\sum_{k}\delta(b-b_{k})\delta(d-d_{k}).
\label{def:Xi}
\end{equation}
Here, the $(b_{k},d_{k})$ are determined by the PD of the union of balls whose 
radius is $r_i(\alpha)=\sqrt{\alpha}$.
The control parameter in the system is 
the number density $\rho=N/L^D$.

Set $m$ an integer with $m\ge 2$.
Because of uniform configuration,
$Q_{\rm rnd}^{mN,\sqrt[D]{m}L}$ shows ``similar" behavior as
$m$ samples of $Q_{\rm rnd}^{N,L}$
for sufficiently large $N$, where we can neglect any boundary effect.
Therefore, the distribution ${\Xi}_n$ is simply multiplied by $m$ 
\begin{equation}
{\Xi}_n\left(b,d;Q_{\rm rnd}^{mN,\sqrt[D]{m}L}\right)
= m {\Xi}_n\left(b,d;Q_{\rm rnd}^{N,L}\right)
+o(m)
\label{3}.
\end{equation}
We expect that Eq.~(\ref{3}) be satisfied even when $m$ is real number.
The scaling in Eq.~(\ref{3}) leads to
\begin{equation}
{\Xi}_n\left(b,d:Q_{\rm rnd}^{N,L}\right)
=N \tilde f(b,d)+o(N).
\end{equation}
Here $\tilde f(b,d)$ is the mean value of ${\Xi}_n/N$ with respect to the
uniform random distribution and is independent of $N$ and $L$,
for the sufficiently large $N$ due to the central limit theorem.

In particular, for the uniform random configuration,
the birth and death scales change 
as $(b_{k},d_{k})\to (\alpha^2b_{k},\alpha^2 d_{k})$
under the transformation,
$Q_{\rm rnd}^{N,L}\to 
\alpha Q_{\rm rnd}^{N,L}\equiv (\alpha\vec x_1,\alpha\vec x_2,...,\alpha\vec x_N)$.
In addition, 
$\alpha Q_{\rm rnd}^{N,L}$ is identical to $Q_{\rm rnd}^{N,\alpha L}$.
Therefore
\begin{equation}
{\Xi}_n\left(b,d;Q_{\rm rnd}^{N,\alpha L}\right)
=\frac{1}{\alpha^4}{\Xi}_n\left(
\frac{b}{\alpha^2},\frac{d}{\alpha^2};Q_{\rm rnd}^{N,L}\right)\label{1}.
\end{equation}

By using Eqs.~(\ref{3}) and (\ref{1}), we obtain 
\begin{equation}
{\Xi}_n\left(b,d;Q_{\rm rnd}^{N,L}\right)
=\frac{v^4}{u^{1+4/D}}
{\Xi}_n\left(
 \frac{b}{(\sqrt[D]{u}/v)^2}, \frac{d}{(\sqrt[D]{u}/v)^2};Q_{\rm rnd}^{uN,vL}\right),
\end{equation}
for the arbitrary $u,v$.
Then, by setting $u=1/N$, $v=1/L$, 
\begin{equation}
{\Xi}_n\left(b,d;Q_{\rm rnd}^{N,L}\right)
=N\rho^{4/D}
{\Xi}_n\left(\rho^{2/D}b, \rho^{2/D}d;Q_{\rm rnd}^{1,1}\right)
+o(N)
\label{scaling:equation}
\end{equation}
is obtained.
Here, the argument $Q^{1,1}_{\rm rnd}$ is formal notation.
It should be rewritten by using Eq.~(\ref{3}) as
\begin{equation}
{\Xi}_n\left(b,d;Q_{\rm rnd}^{N,L}\right)
=N\rho^{4/D}
f\left(\rho^{2/D}b, \rho^{2/D}d\right)+o(N),
\end{equation}
where $\tilde f=\rho^{4/D}f$ and $f$ is dimensionless function.
The bottom panels in Fig.~\ref{scaling:random}
shows the collapse of $f(\rho^{2/D}b,\rho^{2/D}d)$ for $D=3$.

\end{document}